# Verification of a Universal Operator Growth Hypothesis

M. Engelsberg and Wilson Barros Jr.

## Abstract

$F^{19}$ nuclear magnetic resonance free induction decay (FID) data are used to verify the predictions of a universal growth hypothesis for the Lanczos coefficients proposed by Parker et al. Our results, based upon a fit of the data over a wide range of values, strongly support this hypothesis and permit to calculate the growth parameter $\alpha$ for three crystal orientations. For the magnetic field parallel the [100] crystal axis, for example, we found $\alpha = 3.161 \times 10^4 \, sec^{-1}$. The special experimental conditions required for the observability of a singularity in the analytic continuation of the FID, which from the experimental data was found to be of branch-point type, are discussed.

## I. INTRODUCTION

The nuclear magnetic resonance free induction decays (FID`s) in $CaF_2$ constitute almost ideal experimental data to validate quantum dynamical aspects in Hamiltonian systems. Since $F^{19}$ , with 100% natural abundance, has spin I = ½ the interactions are purely dipole-dipole without any quadrupolar broadening effects. Furthermore the $F^{19}$ spins form a simple rigid cubic lattice with no motional effects at room temperature or below. Moreover, among the calcium isotopes $Ca^{40}$ has spin I = 0 and $Ca^{43}$, the only isotope with non-zero spin, is only 0.13% abundant.

Given that nuclear magnetic moments are exceedingly small, Zeeman energies in typical laboratory magnetic fields are much smaller than $kT$ even at liquid Helium temperatures. Thus, one is dealing, to an excellent approximation, with a simple cubic lattice of spin ½ $F^{19}$nuclei interacting via the truncated (1) dipolar Hamiltonian in an infinite temperature regime. Given these unique characteristics, many theoretical investigations have referred to $CaF_2$ NMR FID data as a reliable benchmark to ascertain the correctness of their results (2-8).

Among reported FID data in $CaF_2$ those of Ref. (9) span two orders of magnitude of signal amplitude. Furthermore, the positions of the first eight zeros of the FID are listed for a magnetic field along the [100] crystal direction. These data, although obtained many years ago, contain an experimental verification of a now-recognized universal hypothesis.

Although more recent advances in instrumentation could extend this range by two orders of magnitude (10), some asymptotic features, like the exponential decay reported in Ref (9) and considered to be of a universal nature (6), were confirmed. For the present investigation, the range spanned by the data in Ref (9) was found to be adequate and we employed them throughout this paper.

## I. ANALYTIC BEHAVIOR OF THE FID IN THE COMPLEX PLANE

The FID for a cubic lattice of spin $I = ½$ nuclei in the infinite temperature approximation, represented by $Tr\left(e^{i\mathcal{H}t}I_x e^{-i\mathcal{H}t}I_x\right)$, where $\mathcal{H}$ is the truncated dipolar Hamiltonian, constitutes a classical testing ground for quantum dynamics in Hamiltonian systems. As it will become apparent in the coming sections one important question is whether the analytic continuation of the FID into the complex plane is an entire function, or if it presents singularities. An entire function $f(z)$ is defined by the absence of singularities (poles, or branch points) in the entire complex plane or, equivalently, as a function that can be represented by a power series that converges for all values of the variable thus having an infinite convergence radius $r$. A necessary condition for the FID to be an entire function is that the well-known (1) moment expansion converges for all time.

$$f(t) = \sum_{n=0}^{\infty}(-1)^n M_{2n} t^{2n}/(2n)! \ , \qquad (1)$$

where the FID $f(t)$ is assumed to be an even function of time and $M_{2n}$ are the moments.

The moment expansion of Eq.1 can be shown to be convergent for all values of t if the Carleman (11) series of Eq.2 below diverges,

$$\sum_{n=1}^{\infty}\sqrt{M_2}/M_{2n}^{1/2n} = \infty \ . \qquad (2)$$

Unfortunately, Eq.2 is of little help for the present problem because only a few moments could be calculated exactly (12) or inferred from experimental data (9). All one can do is attempt to check the trend of the few initial terms of the series, although no proof exists that this trend would persist for larger values of n. For example, using the moments $M_2$ to $M_{14}$ for $CaF_2$ with the magnetic field along the [100] crystal axis (9) one obtains the following first seven terms of the Carleman series:

1+0.830+0.740+0.678+0.630+0.588+0.551+……

which indeed decay very slowly. If, for example, one compares the ratio of the seventh to the sixth terms the value (0.551/0.588) = 0.937 is obtained, whereas for the divergent harmonic series $\sum_n 1/n$ the ratio of the seventh to the sixth terms is only 0.857. Despite this trend, it must be emphasized that no rigorous proof of the convergence the series of Eq.1 for all time, or of its lack of convergence, appear to be available.

Given the practical impossibility of obtaining exact higher moments from the dipolar Hamiltonian as the calculations become colossally complex, a radically different approach to the problem proposed by Parker et al. (2) becomes highly attractive. The basic assumption appears to rest on a physically appealing concept connected to the growth of "operator complexity" in Hamiltonian dynamics. It involves the Lanczos coefficients $b_n$, which emerge when the Gram-Schmidt process is applied to orthogonalize a basis of eigenstates of Liouville operators and leads to a continued fraction representation for the Green´s function of the spin auto-correlation (13,14,2). According the to this operator growth hypothesis the Lanczos coefficients, which can be computed from the moments $M_{2n}$, must attain asymptotically their maximum permitted growth which turns out to be linear in $n$:

$$b_n = \alpha n + \gamma + o(1) \ , \qquad (3)$$

where $\alpha > 0$ and $\gamma$ are real numbers. $\alpha\ (sec^{-1})$, referred to as the growth factor, quantitatively captures the growth of operator complexity.

This hypothesis yields asymptotic values of $M_{2n}$ which make the series of Eq.2 convergent and consequently the moment expansion acquires a finite convergence radius $r$ due to singularities of $f(z)$ at $z = \pm ir$. Thus, the FID would not be an entire function according to this theory. This conclusion, although based on a hypothesis, appears to rest on rather solid physical foundations. Some of its consequences could be confirmed from an analysis of our experimental data. Remarkably, values of $\alpha$ and $r = \pi/2\alpha$ for three different orientations of the magnetic field could be determined, as explained in SECTION III.

## II. HADAMARD FACTORIZATION THEOREM AND ZEROS OF THE FID

One of the most prominent features of the FID´s in $CaF_2$ are its zeros, first observed by Lowe and Norberg (15). Since the famous work of Leonhard Euler showing that the function $sin(\pi x)/\pi x$ could be represented by an infinite product of factors $\left(1 - \frac{x}{n}\right)$, where $n = \pm 1, \pm 2, \pm 3$ ... ..are its zeros, it took more than a century to find a generalization of the fundamental theorem of algebra that could allow a representation of a wide class of functions from a knowledge of its zeros. This was accomplished by J. S. Hadamard´s factorization theorem which holds for entire functions of finite order.

Although our aim is to present evidence that shows that the fid cannot be an entire function of the complex variable $z$ we will need Hadamard´s factorization theorem (16) for the discussions of Sections III and IV.

For functions $f(z)$ with no roots at $z = 0$ the function can be represented by:

$$f(z) = e^{P(z)} \prod_n E_p\left(\frac{z}{a_n}\right) \ , \qquad (4)$$

where $a_n$ are all the zeros of $f(z)$ , both real and complex, and $E_p$ is the so-called elementary factor that vanishes at $z = a_n$ and, at the same time, insures that for arbitrary $z$, each factor in the product approaches unity for large enough $n$. $P(z)$ is a polynomial of finite degree $q$ whereas $p$ depends on the growth-rate of the sequence $\{a_n\}$ and is defined as smallest non-negative integer such that the series:

$$\sum_n \frac{1}{|a_n|^{p+1}} \ , \qquad (5)$$

converges.

From the experimental data of Ref.9 for a magnetic field along the [100] axis it was possible to obtain the positions of the first eight real zeros, which were found to only deviate slightly from periodicity. We then conclude that the minimum integer for convergence of Eq.5 should be $p = 1$, which yields elementary factors given by: $E_1\left(\frac{z}{a_n}\right) = (1 - \frac{z}{a_n}) exp\left(\frac{z}{a_n}\right)$ . Moreover, since the FID should be an even function the exponential factors in $E_1$ cancel and Eq.4 reduces to:

$$f(z) = e^{P(z)} \prod_{n=1}^{\infty}\left(1 - \frac{z^2}{a_n^2}\right) \qquad (6)$$

III. ANALYSIS OF THE EXPERIMENTAL FID´s

The aim in Ref (9) was to produce a function that could be adjusted to the best possible fit to the data over the wider possible range, thus permitting to obtain moments $M_{2n}$ from the even order derivatives at $t = 0$. To that end Eq.6 was employed to better fit the behavior near the zeros. However, noticing that the exponential function in front of the product appeared to be gaussian for short times and exponential for long times, instead of using a polynomial in the exponent, as required by Hadamard´s representation for an entire function of finite order, with no other zeros except those on the real axis, a different approach was followed. An empirical function with two adjustable parameters was used. Although other fitting functions were tried, one special function proved to be extremely successful yielding an excellent fit of the data for almost two orders of magnitude of variation. Values of moments $M_{2n}$ for $n = 1,2 \ldots .7$ could be determined which showed excellent agreement with theoretical values (12) of $M_2$, $M_4$, $M_6$, $M_8$ , the only moments which appear to have been calculated so far.

The function employed to fit the data is shown in Eq.7.

$$f(t) = exp\left[CA\left\{1 - \sqrt{1 + \left(\frac{t}{A}\right)^2}\right\}\right] \prod_{n=1}^{\infty}\left(1 - \frac{t^2}{a_n{}^2}\right) \qquad (7)$$

The values of the parameters C and A for the three orientations of the magnetic field are presented in Table I and the details of the calculation of the infinite product function are explained in Ref 11. Given that the zeroes are almost uniformly spaced, the product function approaches $\sin\,(2\pi t/T)/(2\pi t/T)$, where $T$ denotes the average period. For long values of $t$ this contributes an additional $1/t$ factor to the damping caused by the exponential term. It should be pointed out however, that the product function plays a very important role in determining the moments (11). For $\sqrt{M_2}$ and the magnetic field along $[100]$ the calculated contribution of the product term amounts to $79\%$.

Although the function in front of the infinite product in Eq.7 was chosen empirically to achieve an excellent agreement with the experimental FID's, it already displays a remarkable feature. It belongs to the class of functions required by the theory of Ref. 2, as the square root $\sqrt{1 + z^2}$ displays branch-point singularities at $z = \pm i$. From this agreement we could infer that the radius of convergence $r$ of the moment expansion (Eq.1) would by $r = A(\mu sec)$ given in TABLE I. Furthermore, the growth factor $\alpha$ of Eq.1 should be given by:

$$\alpha = \frac{\pi}{2A} \qquad (8)$$

The branch-point singularities in the FID of Eq.7 imply that the frequency-domain spectrum should decay asymptotically as $exp(-|\omega|A)$ times a power law prefactor (12). These results agree with the prediction of Ref.2 and strongly suggest that the hypothesis of maximum permitted growth of the Lanczos coefficients may be correct. It is rather remarkable that the data and the fitting function of Ref.9, that have been available for many years, so directly yield values of the growth parameter of a theory that has only recently become well-recognized.

In addition to Eq.7, a different type of fitting function G(t) was tried in Ref.9. but did not provide nearly as good an agreement as Eq.7. It behaves as an exponentially decaying oscillation for long times and approaches a Gaussian for short time, albeit only up to order $t^2$.

$$G(t) = \left(\frac{a}{b}\right)\left(\frac{\sin(bt)}{\sinh(at)}\right) \qquad (9)$$

G(t) of Eq.9 is not an entire function but, unlike the branch-point singularity of $f(t)$ of Eq.7, it exhibits simple poles for $z = n\pi i/a$ , $n = 1,2,3 \ldots .$ where the denominator vanishes.

Despite the success of Eq.7 to represent the FID over close to two orders of magnitude of data, some deviations become apparent at very long time. This emphasizes that Eq.7 is still only an approximation to the true FID and that smaller terms that become important at long time may still be missing.

TABLE I

| Direction of magnetic field | $A$ $(\mu sec)$ | $C$ $(\mu sec^{-1})$ |
|---|---|---|
| [100] | $49.69 \pm 0.84$ | $0.0462 \pm 0.0005$ |
| [111] | $105.9 \pm 1.6$ | $0.0281 \pm 0.003$ |
| [110] | $115.0 \pm 4.0$ | $0.0425 \pm 0.001$ |

From the data of table I one observes a striking inversion between the positions of the singularities for the orientations [110] and [111]. If, for simplicity, one considers nearest neighbors interactions only and adopts the values of the root second moments as a measure of the strength of the interactions, the largest value of $\alpha$ (smallest value of A) occurs for the [100] orientation, with A[100] = 49.69 $\mu sec$, which corresponds to the strongest dipolar interaction. However for the [110] orientation, which is the next strongest orientation and has a strength 48.5% higher than for the orientation [111], the position of the singularity A[110]=115 µsec can be seen to occur at a larger time than A[111]=105.9 µsec.

The reason for this inversion may be related to the different degrees of spatial connectivity of the spin-spin interactions. Considering interactions of a spin with its six nearest neighbors, 81% of the total second moment for the [110] orientation comes from interactions with its two nearest neighbors along a [100] chain. Unlike this 1-d tendency, for a [111] orientation dipole-dipole interactions with the six nearest neighbors vanish. Since the 12 next-nearest neighbors are distributed more isotropically the 3-d character is dominant, and the connectivity is large. Unlike the 3-d case of Eq.3, a sub-linear growth of the Lanczos coefficients for 1-d systems is predicted, which can make the singularity unobservable. Thus, the 1-d tendency for the [110] orientation may explain the inversion observed in TABLE I.

From an experimental point of view, the question arises as to whether it would be possible to detect a singularity in the analytic continuation of the FID even if the function of Eq.7 were not known. In principle one could use Hadamard's factorization theorem for that purpose. One could consider a finite interval $0 \leq t \leq t_c$ of the FID and by a least squares procedure adjust a polynomial so that $\exp[P(t)]$ of Eq.5 fits the data in this interval to within the experimental uncertainty. One could then verify whether this polynomial still represents the data for $t \geq t_c$. If this is not possible, we could infer that the function is not entire and therefore must have singularities in the complex plane.

Since in our case the FID is approximated very well described by the function of Eq.7 which has branch-point singularities, we know beforehand that this procedure should lead to a substantial departure outside $t_c$. The reason is that given the very good signal to ratio, the value of $t_c$ can be chosen sufficiently larger than the convergence radius $r = A$. We conclude that the experimental condition necessary to assert that a decay is not an entire function is that the signal to noise ratio is large enough so that $t_c$ can be chosen sufficiently larger than $r$ but still smaller than $t_n$, the time when the signal drops to the level of the noise. The simulations shown in Fig.1 illustrate this point.

With the instrumentation available in the earliest pulsed NMR experiments (15), for example, it would not have been possible to detect the singularities. It is interesting to notice that In Abragam's (1) classic textbook on nuclear magnetism it is mentioned the strong resemblance of the FID's with the function $f(t) = \exp(-at^2) \times sinbt/bt$ which is clearly an entire function.

## IV. DETECTABILITY OF SINGULARITIES

We will employ the function in front of the product of Eq.7, which has branch-point singularities at $t/A = \pm i$ to answer the following question: To what extent can one detect the presence of a singularity in FID´s using Hadamard's factorization theorem on raw data in digital form. One chooses time intervals $[0,t_c]$ with increasing values of $t_c$ and performs polynomial least-squares fits of the exponent of Eq.7 given by $ln[f(t)/\prod(t)]$, with data points near zeros removed. The fit is stopped when the accuracy of the agreement reaches a certain level, consistent with the experimental uncertainty. Next, one monitors the agreement between $f(t)/\prod(t)$ and $\exp[P(t)]$ ,where $P(t)$ is the polynomial least-squares fit, outside the interval $[0, t_c]$ . If it persists, we can conclude, according to Hadamard's representation that the function is entire. A departure would indicate the presence of a singularity in the complex plane.

In Figs. 1 we perform this test using as an example the actual function of Eq.7. We plot the ratio $\rho(\tau) = [\exp(P(\tau))]/[f(\tau)/\prod(\tau)]$ as a function of $\tau = t/A$, where $A = 49.69\,\mu sec$ as listed in Table I for [100] parallel to the magnetic field. We adopted the experimental conditions prevailing for the data of Ref.9 . An experimental uncertainty of 2% and a value $\tau_n = t_n/A = 2.66$ for the point where the signal amplitude reaches approximately the level of noise. We used various $\tau_c = t_c/A$ intervals to perform the Hadamard test, both above and below $\tau = 1$ , where the branch-point singularity is located. Within these intervals we performed least-squares fits using Legendre polynomials as an orthonormal basis.

All polynomial fits can be seen from Fig. 1 to depart from $\rho$ =1 for $\tau > \tau_c$.. The departures of $\rho(t)$ from unity above $\tau_c$ are smooth for $\tau < 1$ but become abrupt for $\tau > 1$, above the singularity. One concludes that with $\tau_n = 2.66$ there exists a sizable margin that permits the detection of the singularity. On the other hand, if the singularity were closer to $\tau_n$, for example with $\tau_n \approx 1.2$ , the detection would be difficult.

This suggests that, in a general case, the test could detect whether an FID is an entire function or not using Hadamard's theorem, even if the function used in Eq.2 were not known. However, this can only be true for certain class of functions, namely those where the only zeros of the FID are those measurable on the real axis.

An example of an entire function with roots only on imaginary axis is the complementary error function erfc(t). It is possible to construct an entire function using erfc(t) that approaches a gaussian for short times and an exponential for long times, as required for the data of Ref.9, but the gaussian is only approximate and could not yield as a good fit for short times as Eq.7.

It would be revealing if singularities in $f(z)$ could also be detected in a quasi-1-d system (2), such as the $F^{19}$ FID in fluorapatite (17, 5), as well as in other relevant systems (18,4).

## CONCLUSIONS

We conclude that a particular function with a branch-point singularity in its analytic continuation provides an excellent description of the FID in $CaF_2$ strongly supporting the hypothesis of Parker et al. (2). From the measured positions of the singularities in the time domain for three crystal orientations with respect to the magnetic field it was possible to determine accurate values of the growth parameters $\alpha$ of the theory. Since the Hamiltonian of the system is completely determined and the connectivity of the spin-spin interactions changes with the direction of the magnetic field and affects the values of the growth parameter, the measured values of $\alpha$ could be of significance for understanding the roles of the strength of the interaction and its connectivity in determining the values of $\alpha$

The experimental conditions for the observability of such singularities were discussed and a method for revealing their presence was proposed.

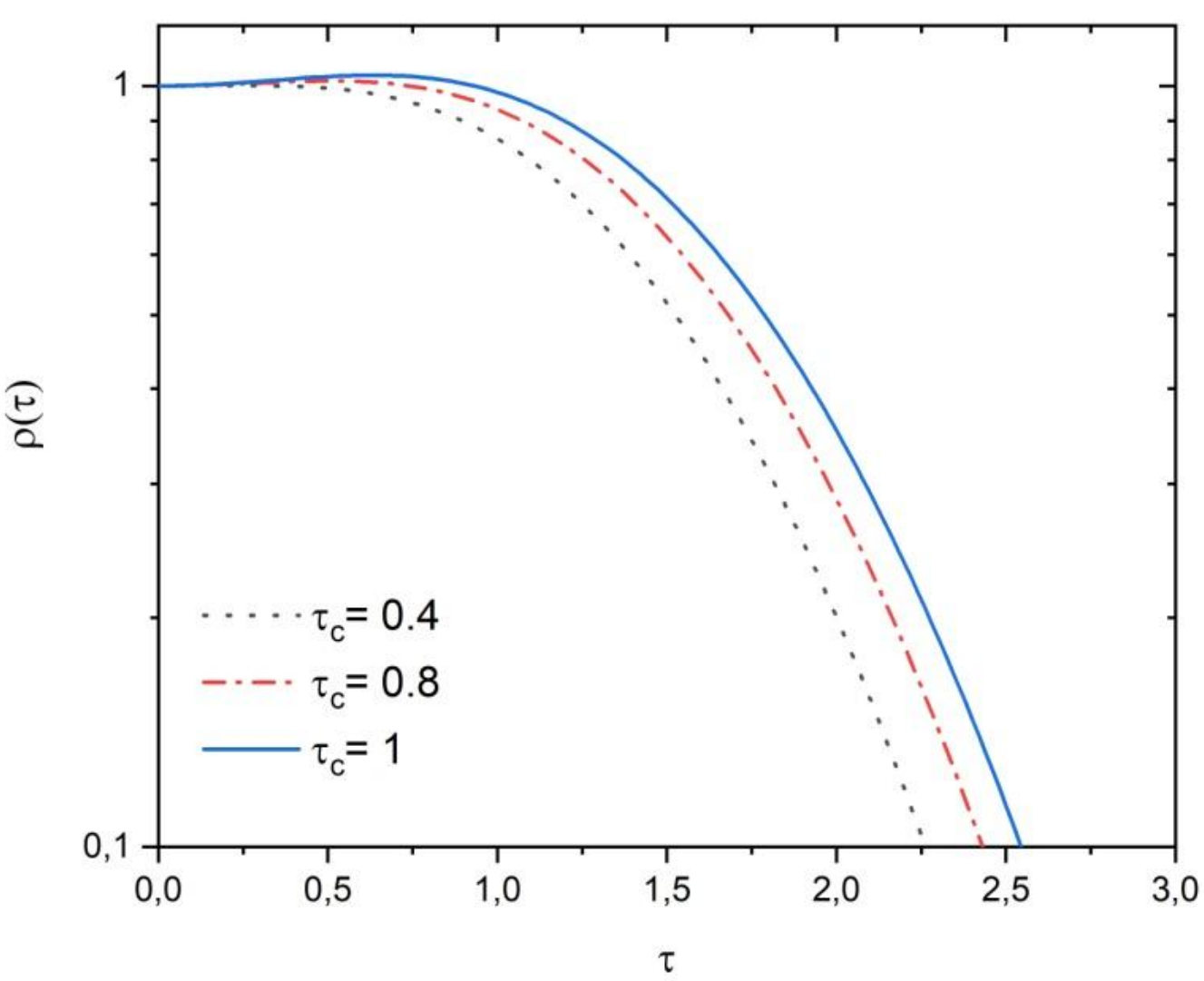

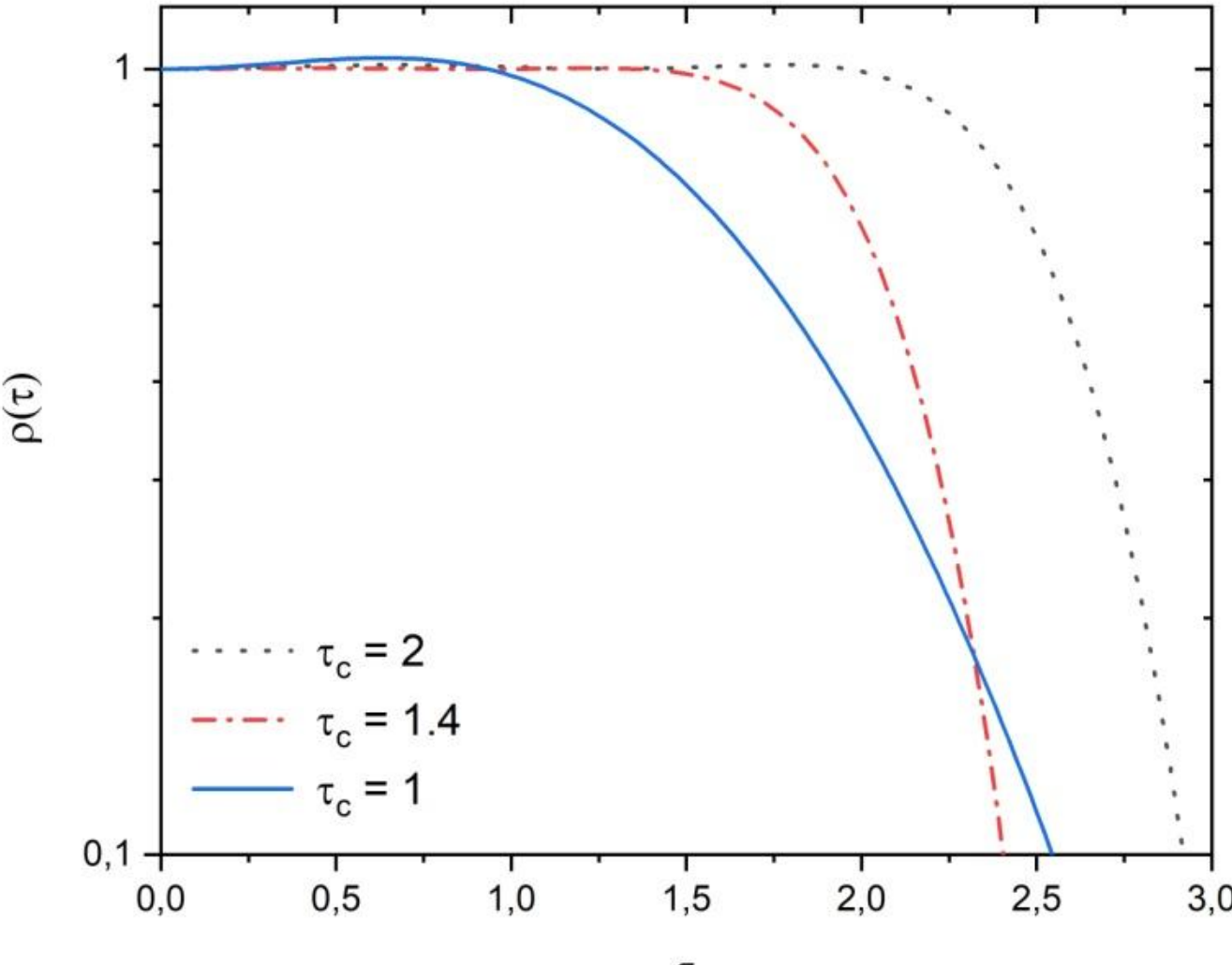

Figure 1 shows plots of $\rho(\tau) = [\exp{(P(\tau))}]/[f(\tau)/\prod(\tau)]$, defined by Eqs.6-7, as a function of $\tau = t/A$ for various values of the polynomial fitting intervals $[0, \tau_c]$. For $\tau_c$= 0.4, 0.8, and 1 (Top) and for $\tau_c$= 1, 1.4, and 2 (bottom). The branch-point singularity is at $\tau = 1$.